\documentclass[preprint,aps,nofootinbib,preprintnumbers,amsmath,amssymb]{revtex4}
\usepackage{graphicx}
\usepackage{dcolumn}
\usepackage{bm}
\usepackage{natbib}
\newcommand{\fft}[2]{{\frac{#1}{#2}}}
\newcommand{\ft}[2]{{\textstyle\frac{#1}{#2}}}
\newcommand{\beq}{\begin{equation}}
\newcommand{\eq}{\end{equation}}
\newcommand{\bea}{\begin{eqnarray}\displaystyle}
\newcommand{\ea}{\end{eqnarray}}

\newcommand{\p}{\partial}

\newcommand{\nn}{\nonumber}

\newcommand{\ctwo}{c_{2I}}

\begin{document}

\preprint{MCTP-08-68}

\title{Black holes in five-dimensional gauged supergravity \\
with higher derivatives}

\author{Sera Cremonini}
\author{Kentaro Hanaki}
\author{James T.~Liu}
\author{Phillip Szepietowski}
\email{{seracre, hanaki, jimliu, pszepiet}@umich.edu}
\affiliation{Michigan Center for Theoretical Physics\\
Randall Laboratory of Physics, The University of Michigan\\
Ann Arbor, MI 48109--1040, USA}

\date{\today}

\begin{abstract}
We examine five-dimensional $\mathcal N=2$ gauged supergravity including
terms up to four derivatives.  These additional terms correspond to the
supersymmetric completion of $R^2$, and were originally obtained in
hep-th/0611329 using conformal supergravity techniques.  Here we integrate
out the auxiliary fields and obtain the on-shell action for minimal
supergravity with such corrections.  We then construct $R$-charged
AdS black holes to linear order in the four derivative terms and investigate
the effect of these corrections on their thermodynamical properties.
Finally, we relate the geometrical coefficients governing
the four-derivative corrections to gauge theory data using
holographic anomaly matching. This enables us to obtain a
microscopic expression for the entropy of the solutions.
\end{abstract}

\maketitle

\section{Introduction}

While one of the major achievements of modern physics has been the development
of fundamental quantum field theories of matter, extending this to quantum
gravity remains a challenge.  In particular, conventional quantization of the
Einstein-Hilbert action leads to a non-renormalizable theory.  Nevertheless,
both gravity and supergravity theories remain viable as effective field
theories describing the low-energy limit of a UV complete theory such as
string theory.  Viewed in this light, it is then natural to explore higher
derivative corrections to the two-derivative action.

Independent of supergravity, many people have considered higher derivative
gravity theories such as $f(R)$ gravity, curvature-square theories, and so
on.  In terms of a derivative expansion, the first non-trivial terms enter
at the $R^2$ level
\begin{equation}
e^{-1}\mathcal L=R+\alpha_1 R^2+\alpha_2 R_{\mu\nu}R^{\mu\nu}
+\alpha_3R_{\mu\nu\rho\sigma}R^{\mu\nu\rho\sigma}+\cdots.
\label{eq:a1a2a3}
\end{equation}
In general, these additional terms modify the graviton propagator and give
rise to ghosts (with the exception of the Gauss-Bonnet combination).  While
this was initially viewed as an argument against higher derivative gravity,
these pathologies only show up at the Planck scale, where traditional
quantum gravity is already ill-defined due to its non-renormalizability.
Furthermore, from the modern effective field theory point of view, such higher
derivative terms are necessarily present, and carry information of the
underlying UV complete theory.

A natural place to explore higher derivative supergravity theories is in
the context of string theory, which gives rise to an effective low energy
supergravity including higher derivative corrections.  For example, it has
been long known from string theory that the first curvature corrections to
the Type II supergravity action appear at $R^4$ order
\cite{Gross:1986iv,Grisaru:1986vi,Freeman:1986zh}, while corrections
to heterotic supergravity first appear at $R^2$ order
\cite{Gross:1986mw,Metsaev:1987zx}.  Of course, even in the absence of
stringy computations, supersymmetry itself puts strong constraints on the
form of the higher derivative terms.  Thus the absence of $R^2$ terms in
Type II supergravity may also be viewed as a consequence of maximal
supersymmetry.  In general, the use of supersymmetry to constrain the form of
the interactions is extremely powerful, and this is simply another example of
this phenomenon.

In this paper, we investigate black holes in higher-derivative corrected
five-dimensional $\mathcal N=2$ gauged supergravity.  Our motivation is
two-fold.  Firstly, we are interested in exploring the nature of stringy
corrections to supergravity and in particular whether such higher-order
corrections may smooth out singular horizons of small black holes.  Secondly,
five-dimensional gauged supergravity is a natural context in which to explore
AdS/CFT, and black holes are important thermal backgrounds for this duality.
By working out these gravity corrections, we may learn more about
finite-coupling as well as $1/N$ effects in the dual $\mathcal N=1$
super-Yang-Mills theory.

Because of the reduced supersymmetry, we expect the first corrections to
$\mathcal N=2$ gauged supergravity to occur at $R^2$ order.  For this reason,
we will limit our focus on four-derivative terms in the effective supergravity
action.  While in principle these terms may be derived directly from string
theory, doing so would involve specific choices of string compactifications
down to five dimensions as well as the potential need to work out contributions
from the Ramond-Ramond sector.  To avoid these issues, we instead make use of
supersymmetry, and in particular the result of \cite{Hanaki:2006pj}, which
worked out the supersymmetric completion of the $A\wedge\mbox{Tr}\,R\wedge R$
term in $\mathcal N=2$ supergravity coupled to an arbitrary number of vector
multiplets using the superconformal tensor calculus methods developed in
\cite{Kugo:2000hn,Bergshoeff:2001hc,Fujita:2001kv,Bergshoeff:2004kh}.

Although we are not aware of an actual uniqueness proof, we expect the
four-derivative terms constructed in \cite{Hanaki:2006pj} to be uniquely
determined by supersymmetry (modulo field redefinitions).  The ungauged
story is rather elegant, and may be tied to M-theory compactified on a
Calabi-Yau three-fold.  In this case the higher derivative corrections
are given by
\begin{equation}
e^{-1}\delta\mathcal L=\ft1{24}c_{2I}\left[
\ft1{16}\epsilon_{\mu\nu\rho\lambda\sigma}A^{I\,\mu}R^{\nu\rho\alpha\beta}
R^{\lambda\sigma}{}_{\alpha\beta}+\cdots\right],
\end{equation}
where the ellipses denote the supersymmetric completion of the $A\wedge
\mbox{Tr}\,R\wedge R$ Chern-Simons term.  Comparing this term with the
Calabi-Yau reduction of the M5-brane anomaly term demonstrates
that the coefficients $c_{2I}$ are related to the second Chern class on the
Calabi-Yau manifold.  The higher-derivative corrected action has recently
been applied to the study of five-dimensional black holes in string theory
(see {\it e.g.}~\cite{Castro:2008ne} and references therein).

While much has already been made of the higher-derivative corrections to
ungauged supergravity, here we are mainly interested in the gauged
supergravity case and resulting applications to AdS/CFT.  In this case, the
natural setup would be to take IIB string theory compactified on
AdS$_5\times Y^5$ where $Y^5$ is Sasaki-Einstein, which is dual to
$\mathcal N=1$ super-Yang-Mills theory in four dimensions.  While the
four-derivative terms worked out in \cite{Hanaki:2006pj} apply equally well
to both gauged and ungauged supergravity, in this case their stringy origin
is less clear.  As we will show, however, the $c_{2I}$ coefficients governing
the four-derivative terms may be related to gauge theory data using
holographic anomaly matching.

Before constructing the $R$-charged black holes in the higher-derivative
corrected theory, we first integrate out the auxiliary fields of the
off-shell formulation, yielding an on-shell supergravity action.  Throughout
this paper, we furthermore work in the truncation to minimal supergravity
involving only the graviton multiplet ($g_{\mu\nu},A_\mu,\psi_\mu$).  While
this on-shell action is implicit in the work of \cite{Hanaki:2006pj}, we
find it useful to have it written out explicitly, as it facilitates
comparison with other recent results.  This is especially of interest in
providing a more rigorous supergravity understanding of the $R^2$ corrections
to shear viscosity \cite{Kats:2007mq,Brigante:2007nu,Brigante:2008gz}
and drag force \cite{Fadafan:2008gb,VazquezPoritz:2008nw}.

The outline of the paper is as follows.
Section II is dedicated to obtaining the on-shell supergravity action.
In Section III we relate the gravitational parameters
$\kappa_5^2$ and $c_{2}$ (the coefficients governing
the four-derivative terms) to the central charges $a$, $c$ of the dual CFT.
In Section IV we construct static
stationary $R$-charged AdS black holes with spherical, flat and hyperbolic
($k=1,0,-1$) horizons.  These solutions, given to linear order in  $c_{2}$,
extend the well-known
black hole solutions of the two-derivative theory \cite{Behrndt:1998jd}.
We also present a brief discussion on the effects of the higher derivative
corrections on the structure of the horizon.
Following this, in Section V we study some basic thermodynamical properties of the
black holes, including their temperature and entropy.
We conclude in Section VI with a discussion.

\section{Higher Derivative Gauged Supergravity}

In this section we investigate five-dimensional $\mathcal N=2$
supergravity with the inclusion of (stringy) higher-derivative corrections.
We are mainly interested in the case of gauged supergravity, which is the
natural setting for the AdS/CFT setup.  Because of the reduced amount of
supersymmetry, we expect the first corrections to this theory to occur at
$R^2$ order.  For this reason, we will limit ourselves to four-derivative
terms in the effective supergravity action.

The conventional on-shell formulation of minimal $\mathcal N=2$ gauged
supergravity is given in terms of the graviton multiplet $(g_{\mu\nu},
A_\mu,\psi_\mu^i)$ where $\psi_\mu^i$ is a symplectic-Majorana spinor with
$i=1,2$ labeling the doublet of SU(2).  The bosonic two-derivative
Lagrangian takes the form
\begin{equation}
e^{-1}\mathcal L_0=-R-\ft14F_{\mu\nu}^2+\ft1{12\sqrt3}
\epsilon^{\mu\nu\rho\lambda\sigma}F_{\mu\nu}F_{\rho\lambda}A_\sigma+12g^2,
\label{eq:l0os}
\end{equation}
where $g$ is the coupling constant of the gauged $R$-symmetry, and
where we have followed
the sign conventions of \cite{Hanaki:2006pj}\footnote{We take
$[\nabla_\mu,\nabla_\nu]v^\sigma= R_{\mu\nu\rho}{}^{\, \sigma} \, v^\rho$ and
$R_{ab} = R_{ac\;\;b}^{\;\;\;\,c}$.}.  We are, of course,
interested in obtaining four-derivative corrections to the above Lagrangian
that are consistent with supersymmetry.  Along with purely gravitational
corrections of the form (\ref{eq:a1a2a3}), other possible four-derivative
terms include $F^4$, mixed $RF^2$ and parity violating ones.  Given the large
number of such terms, it would appear to be a daunting task to work out the
appropriate supersymmetric combinations.  Fortunately, however, it is possible
to make use of manifest supersymmetry in the form of superconformal tensor
calculus to construct supersymmetric $R^2$ terms.  (See
{\it e.g.}~\cite{Mohaupt:2000mj} for a nice review, albeit focusing on
four-dimensional $\mathcal N=2$ supergravity.)

The general idea of the superconformal approach is to develop an off-shell
formulation involving the Weyl multiplet that is locally gauge invariant
under the superconformal group.  The resulting conformal supergravity may
then be broken down to Poincar\'e supergravity by introducing a conformal
compensator in the hypermultiplet sector and introducing expectation
values for some of its fields.  One advantage of this method is that the
off-shell formulation admits a superconformal tensor calculus which enables
one to construct supersymmetric invariants of arbitrary order in curvature.
This is in fact the approach taken in \cite{Hanaki:2006pj}, which
worked out the supersymmetric completion of the $A\wedge\mbox{Tr}\,R\wedge R$
term in $\mathcal N=2$ supergravity coupled to an arbitrary number of vector
multiplets.

The basic construction of \cite{Hanaki:2006pj} involves conformal supergravity
({\it i.e.}~the Weyl multiplet) coupled to a set of $n_V+1$ conformal
vector multiplets and a single compensator hypermultiplet.  The resulting
Lagrangian takes the form
\begin{equation}
\mathcal L=\mathcal L_0+\mathcal L_1=\mathcal L_0^{(V)}+\mathcal L_0^{(H)}
+\mathcal L_1,
\label{eq:l0l1}
\end{equation}
where $\mathcal L_0$ corresponds to the two-derivative terms and $\mathcal L_1$
the four-derivative terms.  We have further broken up $\mathcal L_0$ into
contributions $\mathcal L_0^{(V)}$ from the vector multiplets and
$\mathcal L_0^{(H)}$ from the hypermultiplet.

As formulated in \cite{Hanaki:2006pj}, the full Lagrangian $\mathcal L$
contains a set of auxiliary fields which we wish to eliminate in order to
make direct comparison to the on-shell Lagrangian (\ref{eq:l0os}).  To
do so, we simply integrate out the auxiliary fields using their equations
of motion, and the remainder of this section is devoted to this process.
As an important shortcut, we note that when working to linear order in
the correction terms in $\mathcal L_1$, we only need to substitute in the
lowest order expressions for the auxiliary fields \cite{Argyres:2003tg}.
For this reason, we first examine the two-derivative Lagrangian before
turning to the four-derivative terms contained in $\mathcal L_1$.

\subsection{The leading two-derivative action}
\label{L0section}

We begin with the vector multiplet contribution to the two-derivative
Lagrangian \cite{Hanaki:2006pj}
\bea
\label{vectoraction}
e^{-1} {\cal L}_0^{(V)} &=& {\cal N}(\ft12D-\ft14R+3v^2)
+ 2 {\cal N}_I v^{\mu\nu}F^I_{\mu\nu}+
{\cal N}_{IJ} \ft{1}{4}F^I_{\mu \nu} F^{J\, \mu \nu} +
\ft{1}{24}c_{IJK}\epsilon^{\mu\nu\rho\lambda\sigma}
A^I_\mu F^J_{\nu\rho} F^K_{\lambda\sigma}\nn \\
&&- {\cal N}_{IJ}\bigl(\ft12 {\cal D}^\mu M^I {\cal D}_\mu M^J +
Y^I_{ij} Y^{J\; ij}\bigr)\, ,
\ea
where $M^I$, $A_\mu^I$ and $Y^I_{ij}$ ($I,J=1,2,\ldots,n_v+1$) denote, respectively,
the scalar fields, the gauge fields and the $SU(2)$-triplet
auxiliary fields in the $n_v+1$ vector multiplets.  In addition, the scalar
$D$ and the two-form $v_{\mu\nu}$ are auxiliary fields coming from the Weyl
multiplet.  The prepotential $\mathcal N$ and its functional derivatives are
given by the standard expressions
\begin{equation}
\mathcal N=\ft16 c_{IJK}M^IM^JM^K,\qquad
\mathcal N_I=\ft12 c_{IJK}M^JM^K,\qquad
\mathcal N_{IJ}= c_{IJK}M^K.
\label{eq:prepot}
\end{equation}
For future reference, we also note the useful relations
\begin{equation}
\mathcal N_IM^I=3\mathcal N,\qquad\mathcal N_{IJ}M^J=2\mathcal N_I.
\end{equation}

Turning next to the hypermultiplet Lagrangian, we have \cite{Hanaki:2006pj}
\bea
\label{hyperaction}
e^{-1}
{\cal L}_0^{(H)} &=& 2 \bigl[{\cal D}^\mu {\cal A}^{\bar{\alpha}}_i  {\cal D}_\mu {\cal A}_{\alpha}^i
+ {\cal A}^{\bar{\alpha}}_i \,(g\, M)^2 \, {\cal A}_{\alpha}^i
+ 2g Y^{ij}_{\alpha\beta}  {\cal A}^{\bar{\alpha}}_i {\cal A}^{\beta}_j \bigr] +
{\cal A}^2 (\ft14D+\ft38R-\ft12v^2) \, .
\ea
In general, ${\cal A}_{\alpha}^i$ are a set of $4\times n_H$ hypermatter scalars
carrying both the SU(2) index $i$ and the index $\alpha=1,2,\ldots,2n_H$ of
USp($2n_H$).  (We use the SU(2) index raising convention $A^i = \epsilon^{ij}
A_j$ and $A_i = A^j \epsilon_{ji}$ with $\epsilon_{12}=\epsilon^{12}=1$).
Note that we have gauged a subgroup $G$ of USp($2n_H$), so that the covariant
derivative appearing above is given by
\begin{equation}
{\cal D}_\mu {\cal A}^{\alpha}_i = \p_\mu {\cal A}^{\alpha}_i  -g A^I_\mu
t_I{\cal A}^{\alpha}_i+ {\cal A}^{\alpha}_j V_\mu^j{}_{i} \, ,
\end{equation}
where $t_I$ are the generators of the gauge symmetry and where
$V_\mu^{ij}$ is an additional auxiliary field belonging to the Weyl multiplet.
Finally, we have defined $M \equiv M^I t_I$, where $M^I$ are the
vector multiplet scalars.

For simplicity, we focus on a single compensator and choose the conventional
gauging of the diagonal U(1) in the SU(2) $R$-symmetry.  In this case,
the action of $M$ on the hyperscalars is given by
\beq
\label{Mexp}
M\mathcal A^\alpha_i
=M^I t_I {\cal A}^\alpha_i
= M^I P_I (i \sigma^3)^\alpha_\beta {\cal A}^\beta_i \, ,
\eq
while the covariant derivative becomes
\begin{equation}
{\cal D}_\mu {\cal A}^{\alpha}_i = \p_\mu {\cal A}^{\alpha}_i-g A^I_\mu
P_I (i \sigma^3)^\alpha_\beta
{\cal A}^{\beta}_i+ {\cal A}^{\alpha}_j V_\mu^j{}_{i} \, .
\end{equation}
Here $P_I$ denote the charges associated with the gauging.
Furthermore, ${\cal A}^2 \equiv  {\cal A}^{\bar{\alpha}}_i {\cal A}_{\alpha}^i
={\cal A}^{\beta}_i d_\beta^{\;\;\alpha} {\cal A}^i_{\alpha}$, where the metric $d_\beta^{\;\;\alpha}$
is arranged to be a delta function as appropriate for a compensator
\cite{Hanaki:2006pj}.

Combining (\ref{vectoraction}) with (\ref{hyperaction}), the complete
two-derivative action is given by
\begin{eqnarray}
\label{L0totalA}
e^{-1}
{\cal L}_0 &=&
\ft{1}{4} D (2{\cal N}+ {\cal A}^2) + R \left(\ft{3}{8}\,{\cal A}^2
-\ft14\mathcal N\right)
+ v^2 (3{\cal N}-\ft12{\cal A}^2) \nn \\
&&+ 2 {\cal N}_I v^{\mu\nu}F^I_{\mu\nu}
+ {\cal N}_{IJ} ( \ft{1}{4}F^I_{\mu \nu} F^{J\; \mu \nu}
- \ft12 {\cal D}^\mu M^I {\cal D}_\mu M^J )
+\ft{1}{24}c_{IJK}\epsilon^{\mu\nu\rho\lambda\sigma}
A^I_\mu F^J_{\nu\rho} F^K_{\lambda\sigma}\nn \\
&&- {\cal N}_{IJ} Y^I_{ij} Y^{J\; ij} + 2 \bigl[{\cal D}^\mu {\cal A}^{\bar{\alpha}}_i  {\cal D}_\mu {\cal A}_{\alpha}^i
+ {\cal A}^{\bar{\alpha}}_i \,(g\, M)^2 \, {\cal A}_{\alpha}^i
+ 2g Y^{ij}_{\alpha\beta}  {\cal A}^{\bar{\alpha}}_i {\cal A}^{\beta}_j \bigr]  \, .
\end{eqnarray}
At the two-derivative level, the auxiliary field $D$ plays the role of a
Lagrange multiplier, yielding the constraint
\beq
2{\cal N}+ {\cal A}^2=0 \, .
\eq
Thus we can recover the standard very special geometry constraint ${\cal N}=1$
by setting ${\cal A}^2=-2$.  (This fixing of the dilatational gauge
transformation is in fact the purpose of the conformal compensator).
This then brings the Lagrangian to the following form:
\bea
\label{L0total}
{\cal L}_0 &=&
\ft{1}{2} D ({\cal N}-1) -\ft14R ({\cal N}+3) + v^2 (3{\cal N} +1 ) +
2 {\cal N}_I v^{\mu\nu}F^I_{\mu\nu}\nn \\
&&+ {\cal N}_{IJ} ( \ft{1}{4}F^I_{\mu \nu} F^{J\, \mu \nu}
- \ft12 {\cal D}^\mu M^I {\cal D}_\mu M^J )
+\ft{1}{24}\,c_{IJK}\epsilon^{\mu\nu\rho\lambda\sigma}
A^I_\mu F^J_{\nu\rho} F^K_{\lambda\sigma}\nn \\
&&- {\cal N}_{IJ} Y^I_{ij} Y^{J\; ij} + 2 \bigl[{\cal D}^\mu {\cal A}^{\bar{\alpha}}_i  {\cal D}_\mu {\cal A}_{\alpha}^i
+ {\cal A}^{\bar{\alpha}}_i \,(g\, M)^2 \, {\cal A}_{\alpha}^i
+ 2g Y^{ij}_{\alpha\beta}  {\cal A}^{\bar{\alpha}}_i {\cal A}^{\beta}_j \bigr]  \, .
\ea
\subsubsection{Integrating out the auxiliary fields}

The action (\ref{L0total}) can be written in a more familiar on-shell form
by integrating out the auxiliary fields.  We will do this in two steps by
first eliminating the fields $\mathcal A_i^\alpha$, $V_\mu^{ij}$ and
$Y_{ij}^I$ and then eliminating $D$ and $v_{\mu\nu}$.

We start by fixing the SU(2)
symmetry by taking ${\cal A}^\alpha_i = \delta^\alpha_i$, which identifies
the indices in the hypermultiplet scalar.  The equation of motion for
$V_\mu^{ij}$ is then given by
\beq
\label{Veom}
V_\mu^{ij} = gP_I (i\sigma^3)^{ij} A^I_\mu \, ,
\eq
which also results in ${\cal D}_\mu {\cal A}^{\alpha}_i =0$.  Turning next
to $Y^I_{ij}$, we first note that
\beq
\label{Yexp}
Y^{ij}_{\alpha \beta} \, {\cal A}^{\bar{\alpha}}_i {\cal A}^\beta_j
= Y^{I\; ij}P_I \, (i\sigma^3)_{ij} \, .
\eq
Varying (\ref{L0total}) with respect to $Y^I_{ij}$ then gives us the equation
of motion
\beq
\label{Yaux}
Y^I_{ij}=2({\cal N}^{-1})^{IJ} P_J (i\sigma^3)_{ij} \, .
\eq

Using the above to eliminate $\mathcal A_i^\alpha$, $V_\mu^{ij}$ and
$Y_{ij}^I$ from the two-derivative action (\ref{L0total}), we end up with
\bea
\label{L0gauged}
e^{-1} {\cal L}_0 &=&  \ft12 D ({\cal N}-1)-\ft14{R}({\cal N}+3)
+v^2(3{\cal N}+1)+ 2 {\cal N}_I v^{\mu\nu} F^I_{\mu\nu}\nn \\
&+& {\cal N}_{IJ} \bigl( \ft{1}{4} F^I_{\mu \nu} F^{J\, \mu \nu}
-\ft12 \partial^\mu M^I \partial_\mu M^J\bigr)+
\ft{1}{24}c_{IJK}\epsilon^{\mu\nu\rho\lambda\sigma}
A^I_\mu F^J_{\nu\rho} F^K_{\lambda\sigma}\nn\\
&+&  8g^2 ({\cal N}^{-1})^{IJ} P_I P_J  + 4g^2 (P_I M^I)^2   \, ,
\ea
where the last line corresponds to the gauged supergravity potential
\begin{equation}
\label{scalarpotential}
V=-4g^2[2 ({\cal N}^{-1})^{IJ} P_I P_J + (P_I M^I)^2] \, .
\end{equation}
Note that, with abelian gauging, the covariant derivative acts trivially
on the vector multiplet scalars, $\mathcal D_\mu M^I=\partial_\mu M^I$.

To remove the remaining auxiliary fields $D$ and $v_{\mu\nu}$ from
(\ref{L0gauged}) we must turn to the equations of motion for this system.
Varying the action with respect to $D$, $v_{\mu\nu}$, $M^I$ and $A_{\mu}^I$
yields, respectively,
\bea
0&=&\ft12(\mathcal N-1)\, , \\
\label{vaux}
0&=&2(3\mathcal N+1)v_{\mu\nu}+2\mathcal N_IF_{\mu\nu}^I\, , \\
\label{gaugedScalars}
0 &=& \ft12\mathcal N_I(D-\ft12R+6v_{\mu\nu}v^{\mu\nu})+2\, {\cal N}_{IJ} F^J_{\mu\nu} v^{\mu\nu} +
\ft14 \, c_{IJK} \, F_{\mu\nu}^J F^{K\; \mu\nu} + {\cal N}_{IJ} \Box M^J \nn \\
&& + \ft12 \, c_{IJK} \, \p_\mu M^J \p^\mu M^K
-g^2\frac{\delta V}{\delta M^I}\, , \\
0 &=&-\nabla^\nu[4\mathcal N_Iv_{\nu\mu}+\mathcal N_{IJ}F^J_{\nu\mu}]
+\ft18C_{IJK}\epsilon_\mu{}^{\nu\rho\lambda\sigma}
F^J_{\nu\rho}F^K_{\lambda\sigma}\, .
\ea
In addition, the Einstein equation is given by:
\bea
0 &=& \ft14(\mathcal N+3)(R_{\mu\nu}-\ft12g_{\mu\nu}R)+\ft14(\mathcal N-1)
Dg_{\mu\nu}-\ft14(\nabla_\mu\nabla_\nu\mathcal N-g_{\mu\nu}\Box\mathcal N)
\nonumber\\
&& + \ft12\mathcal N_{IJ}(\partial_\mu M^I\partial_\nu M^J-\ft12g_{\mu\nu}
\partial_\lambda M^I\partial^\lambda M^J)
-2(3\mathcal N+1)(v_{\mu\lambda}v_\nu{}^\lambda
-\ft14g_{\mu\nu}v_{\lambda\sigma}v^{\lambda\sigma})\nonumber\\
&& -4\mathcal N_I(F^I_{(\mu}{}^\lambda v_{\nu)\lambda}
-\ft14g_{\mu\nu}F^I_{\lambda\sigma}v^{\lambda\sigma})
-\ft12\mathcal N_{IJ}(F^I_{\mu\lambda}F^J_\nu{}^\lambda
-\ft14g_{\mu\nu}F^I_{\lambda\sigma}F^{J\,\lambda\sigma}) -\ft12g_{\mu\nu}V\, .
\label{eq:eom0}
\ea
We are now in a position to start solving for the auxiliary fields $D$ and
$v_{\mu\nu}$.  Inserting the very special geometry constraint ${\cal N}=1$
(enforced by the equation of motion for $D$) into (\ref{vaux}) yields
\begin{equation}
v_{\mu\nu}=-\ft14\mathcal N_IF^I_{\mu\nu}\,.
\label{eq:vab}
\end{equation}
We may now eliminate $\mathcal N$ and $v_{\mu\nu}$ from the lowest order
Maxwell and Einstein equations to obtain
\begin{eqnarray}
\nabla^\nu[(\mathcal N_I\mathcal N_J-\mathcal N_{IJ})F_{\nu\mu}^J]
&=&-\ft18C_{IJK}\epsilon_\mu{}^{\nu\rho\lambda\sigma}F^J_{\nu\rho}
F^K_{\lambda\sigma},\nonumber\\
R_{\mu\nu}-\ft12g_{\mu\nu}R&=& -\ft12\mathcal N_{IJ}(\partial_\mu M^I
\partial_\nu M^J-\ft12g_{\mu\nu}\partial_\lambda M^I\partial^\lambda M^J)
\nonumber\\
&&-\ft12(\mathcal N_I\mathcal N_J-\mathcal N_{IJ})(F^I_{\mu\lambda}
F^J_\nu{}^\lambda-\ft14g_{\mu\nu}F^I_{\lambda\sigma}F^{J\,\lambda\sigma})
+\ft12g_{\mu\nu}V \, .
\label{MaxwellEinstein}
\end{eqnarray}

Turning next to the scalar equations of motion, we note that the $n_v+1$
equations may be decomposed into $n_v$ equations for the constrained
scalars $M^I$, along with one equation for the Lagrange multiplier $D$.  To
solve for $D$, we multiply the scalar equation by $M^I$ and obtain:
\bea
D-\ft12R+6v_{\mu\nu}v^{\mu\nu}&=& -\ft83\mathcal N_IF^I_{\mu\nu}v^{\mu\nu}
-\ft16\mathcal N_{IJ}F_{\mu\nu}^IF^{J\,\mu\nu}
-\ft13\mathcal N_{IJ}\partial_\mu M^I\partial^\mu M^J \nn \\
&&-\ft43\mathcal N_I\Box M^I +\ft23 M^I \frac{\delta V}{\delta M^I}.
\label{eq:drv}
\ea
Substituting in $R$ and $v_{\mu\nu}$ then allows us to express
the auxiliary field $D$ entirely in terms of physical fields:
\bea
D &=& -\ft7{12}\mathcal N_{IJ}\partial_\mu M^I\partial^\mu M^J
- \ft43\mathcal N_I
\Box M^I+\ft14(\mathcal N_I\mathcal N_J-\ft12\mathcal N_{IJ})F_{\mu\nu}^I
F^{J\,\mu\nu}
- \ft56 V + \ft23 M^I \frac{\delta V}{\delta M^I} \nn \\
&=& -\ft7{12}\mathcal N_{IJ}\partial_\mu M^I\partial^\mu M^J - \ft43\mathcal N_I
\Box M^I+\ft14(\mathcal N_I\mathcal N_J-\ft12\mathcal N_{IJ})F_{\mu\nu}^I
F^{J\,\mu\nu} \nn \\
&& + 2g^2[6 P_I P_J (\mathcal N^{-1})^{IJ}- P_I P_J M^I M^J]\, .
\label{eq:d0}
\ea
By using (\ref{eq:drv}), the equation of motion for the constrained scalars
(\ref{gaugedScalars}) can be rewritten in the following form:
\begin{eqnarray}
&&\left(\delta_I^J-\frac{{\mathcal N}_I M^J}{3}\right)
\biggl[c_{JKL}(\p_\mu M^K \p^\mu M^L +
2 M^K \Box M^L)\nn\\
&&\kern12em -(\mathcal N_{JK}\mathcal N_L- \ft12 c_{JKL})F^K F^L
-\frac{\delta V}{\delta M^J}  \biggr]=0.
\label{Scalars}
\end{eqnarray}

We now have all the ingredients we need to write down the
on-shell two-derivative Lagrangian:
\bea
e^{-1}\mathcal L &=& -R - \ft12\mathcal N_{IJ}\partial_\mu M^I\partial^\mu M^J
-\ft14(\mathcal N_I\mathcal N_J-\mathcal N_{IJ})F_{\mu\nu}^IF^{J\,\mu\nu} \nn \\
&& +\ft1{24}\, c_{IJK} \, \epsilon^{\mu\nu\rho\lambda\sigma}
A_\mu^IF_{\nu\rho}^JF_{\lambda\sigma}^K + 4g^2[2({\cal N}^{-1})^{IJ}P_IP_J
+(P_I M^I)^2] \, ,
\label{eq:onshell}
\ea
where now the $M^I$ are a set of constrained scalars satisfying the
very special geometry condition $\mathcal N =1$.  The Lagrangian perfectly
matches the bosonic sector of the standard two-derivative $\mathcal N=2$
supergravity action coupled to $n_v$ vector multiplets.  The resulting
equations of motion are given by (\ref{MaxwellEinstein}) and (\ref{Scalars}).

Here, we are mainly concerned with the truncation of
(\ref{eq:onshell}) to the case of pure supergravity.
This is accomplished by setting the scalars to constants and by defining a
single graviphoton $A_\mu$ according to%
\footnote{Note that our definition differs by a factor of $1/3$ from the
conventional one where $A_\mu = A_\mu^I{\cal N}_I$.}
\begin{equation}
M^I=\bar M^I,\qquad A_\mu^I=\bar M^I A_\mu.
\label{eq:0trunc}
\end{equation}
While the constants $\bar M^I$ are arbitrary moduli in the ungauged case,
in the gauged cause they must lie at a critical point of the potential
(\ref{scalarpotential}) given by solving
\begin{equation}
\left(\delta^J_I-\fft{\mathcal N_IM^J}3\right)\fft{\delta V}{\delta M^J}=0.
\end{equation}
By demanding that the critical point is supersymmetric, we find that the
constant scalars satisfy%
\footnote{These expressions can be obtained by making use of the hyperino
and gauging SUSY variations, as well as the equation of motion for the
auxiliary field $Y^I_{ij}$. We refer the reader to \cite{Hanaki:2006pj}
for more details.}:
\beq
\label{globalAds}
P_I \bar M^I = \frac{3}{2}\,,\qquad(\bar{\mathcal N}^{-1})^{IJ} P_I P_J
= \frac{3}{8}\, .
\eq
in which case the potential becomes $\bar V=-12g^2$.
The resulting Lagrangian for the bosonic fields of the supergravity
multiplet $(g_{\mu\nu},A_\mu)$ then reads
\begin{equation}
\label{L0truncation}
e^{-1}\mathcal L=-R-\ft34F_{\mu\nu}^2+\ft14\epsilon^{\mu\nu\rho\lambda\sigma}
A_\mu F_{\nu\rho}F_{\lambda\sigma}+ 12g^2 \, ,
\end{equation}
which reproduces the conventional on-shell supergravity Lagrangian
(\ref{eq:l0os}) once the graviphoton is rescaled according to
$A_\mu\to A_\mu/\sqrt3$.

While this completes the analysis relevant to the leading, two-derivative
action, we note that the expression for $D$ simplifies further in
the case of constant scalars.  Substituting (\ref{eq:0trunc}) and
(\ref{globalAds}) into the expression (\ref{eq:d0}) for $D$ yields the
simple result
\beq
\label{Dexpr}
D=\ft14(\bar{\mathcal N}_I\bar{\mathcal N}_J-\ft12\bar{\mathcal N}_{IJ})
F_{\mu\nu}^IF^{J\,\mu\nu}=\ft32F_{\mu\nu}^2 \, .
\eq
By taking $\mathcal N=1$, we see that this explicit form of $D$ does not
play a role in the leading expression for the two-derivative Lagrangian.
However, it will become relevant in the discussion of higher derivative
corrections, which we turn to next.

\subsection{Higher-derivative corrections in gauged SUGRA}

We now turn to the four-derivative corrections to the action (\ref{eq:l0l1}),
which we parameterize by $\mathcal L_1$.  For convenience, we separate the
contributions to $\mathcal L_1$ present in the ungauged theory
from those coming strictly from the gauging, $\mathcal L_1 =
\mathcal L_1^{\rm ungauged} + \mathcal L_1^{\rm gauged}$.
The two are given by:
\bea
\label{ungauged}
e^{-1}\mathcal L_1^{\rm ungauged}\!\!&=&\!\ft 1{24} c_{2I}\Bigl[
\ft1{16}\epsilon_{\mu\nu\rho\lambda\sigma}A^{I\,\mu}R^{\nu\rho\alpha\beta}
R^{\lambda\sigma}{}_{\alpha\beta}+\ft18M^IC_{\mu\nu\rho\sigma}
C^{\mu\nu\rho\sigma}+\ft1{12}M^ID^2+\ft16F^I_{\mu\nu}v^{\mu\nu}D\nonumber\\
&&\quad - \ft13M^IC_{\mu\nu\rho\sigma}v^{\mu\nu}v^{\rho\sigma} - \ft12F^{I\,\mu\nu}
C_{\mu\nu\rho\sigma}v^{\rho\sigma}+\ft83M^Iv_{\mu\nu}\nabla^\nu\nabla_\rho
v^{\mu\rho}\nonumber\\
&&\quad-\ft{16}9M^Iv^{\mu\rho}v_{\rho\nu}R_\mu^\nu-\ft29M^Iv^2R
+\ft43M^I\nabla^\mu v^{\nu\rho}\nabla_\mu v_{\nu\rho}
+\ft43M^I\nabla^\mu v^{\nu\rho}\nabla_\nu v_{\rho\mu}\nonumber\\
&&\quad-\ft23M^I\epsilon_{\mu\nu\rho\lambda\sigma}v^{\mu\nu}v^{\rho\lambda}\nabla_\delta v^{\sigma\delta}
+\ft23F^{I\,\mu\nu}\epsilon_{\mu\nu\rho\lambda\sigma}v^{\rho\delta}
\nabla_\delta v^{\lambda\sigma}+F^{I\,\mu\nu}\epsilon_{\mu\nu\rho\lambda\sigma}
v^\rho{}_\delta\nabla^\lambda v^{\sigma\delta}\nonumber\\
&&\quad-\ft43F^{I\,\mu\nu}v_{\mu\rho}v^{\rho\lambda}v_{\lambda\nu}
-\ft13F^{I\,\mu\nu}v_{\mu\nu}v^2+4M^Iv_{\mu\nu}v^{\nu\rho}v_{\rho\lambda}
v^{\lambda\mu}-M^I(v^2)^2\Bigr] \, , \\
\label{L1gauged}
e^{-1}
{\cal L}_1^{\rm gauged}\!\! &=&\! \ft 1{24} c_{2I}  \Bigl[-\ft1{12}
\epsilon_{\mu\nu\rho\lambda\sigma} \,  A^{I\, \mu}
R^{\nu\rho \, ij} (U) R^{\lambda\sigma}_{ij}(U)\nn \\
&&\quad - \ft13 \, M^I R^{\mu \nu \, ij} (U) R_{\mu \nu \, ij}(U)
-\ft43 \,Y^I_{ij}v_{\mu\nu} R^{\mu \nu \; ij}(U) \Bigr] \, ,
\ea
where
\beq
R_{\mu \nu}^{\;\;\; i j}(U) = \p_\mu V_\nu^{ij}-V_{\mu k}^{i} V_{\nu}^{kj} - (\mu \leftrightarrow\nu) \, .
\eq
As we can see, the constants $c_{2I}$ parameterize the magnitude of these
contributions.
Notice that the scalar $D$ no longer acts as a Lagrange multiplier, since
it now appears quadratically in $\mathcal L_1$.  In fact, by varying the full
action $\mathcal L=\mathcal L_0 + \mathcal L_1$ with respect
to $D$, with $\mathcal L_0$ as in (\ref{L0gauged}),  we obtain
the modified very special geometry constraint
\beq
\label{Deom}
{\cal N}=1-\frac{\ctwo}{72}(DM^I + F^{I\, \mu \nu} v_{\mu \nu}) \, ,
\eq
which encodes information about how the scalars $M^I$ are affected
by higher-derivative corrections.

\subsubsection{Integrating out the auxiliary fields}

As in the two-derivative case, in order to obtain a Lagrangian written
solely in terms of the physical fields of the theory we need to eliminate
the auxiliary fields $D$, $v_{\mu\nu}$, $V_{\mu\nu}^i$ and $Y^I_{ij}$ from
$\mathcal L=\mathcal L_0 + \mathcal L_1$. In Sec.~\ref{L0section} we solved
for the auxiliary fields by neglecting higher order corrections, and then
integrated them out of the two-derivative action.  It turns out that the
lowest order expressions for the auxiliary fields are sufficient when working
to linear order in the $c_{2I}$ \cite{Argyres:2003tg}.  This allows us to
reuse the results of the previous section for the auxiliary fields, which we
summarize here:
\bea
\label{Veom1}
V_\mu^{ij} &=& gP_I (i\sigma^3)^{ij} A^I_\mu \, ,\\
\label{Yeom1}
Y^I_{ij} &=& 2({\cal N}^{-1})^{IJ} P_J (i\sigma^3)_{ij} \, , \\
\label{veom1}
v_{\mu\nu} &=& -\ft14\mathcal N_IF^I_{\mu\nu} \, , \\
\label{Deom1}
D &=& \ft14(\mathcal N_I\mathcal N_J-\ft12\mathcal N_{IJ})F_{\mu\nu}^I
F^{J\,\mu\nu}\, .
\ea
While it is valid to use these lowest order expressions, it is important
to realize that the scalar fields are modified because of (\ref{Deom}).
This modification leads to additional contributions to the two-derivative
on-shell action (\ref{eq:onshell}), which combines with $\mathcal L_1$ to
yield the complete action at linear order in $c_{2I}$.

In principle, we may work with the full system of supergravity coupled to
$n_V$ vector multiplets.  However, here we focus on the truncation to pure
supergravity, where the scalars $M^I$ are taken to be non-dynamical.  Even so,
they are not entirely trivial.  While at the two-derivative level, we may
simply set them to constants according to (\ref{eq:0trunc}), here we must
allow for the modification (\ref{Deom}) by defining
\begin{equation}
M^I=\bar M^I+c_2\hat M^I,\qquad A_\mu^I=\bar M^I A_\mu, \qquad c_2\equiv c_{2I}\bar M^I \, ,
\label{eq:sans1}
\end{equation}
where $\hat M^I$ are possible scalar fluctuations that enter at
$\mathcal O(c_2)$.  Substituting this into the expressions (\ref{veom1})
and (\ref{Deom1}) for the auxiliary fields then yields
\beq
v_{\mu\nu} =-\ft34F_{\mu\nu}+\mathcal O(c_2),\qquad
D=\ft32 F^2 + \mathcal O(c_2) \, ,
\label{eq:vlo}
\eq
which match the lowest order expressions for constant scalars.
The modified very special geometry constraint (\ref{Deom}) can now be
simplified further, and becomes
\begin{equation}
\label{finalN}
\mathcal N=1-\fft{c_2}{96}F^2+\mathcal O(c_2^2).
\end{equation}
In general, a solution to the fluctuating scalars $\hat M^I$ ought to come
from the equations of motion.  However, as a shortcut, we make the ansatz
that $\hat M^I$ is proportional to $\bar M^I$.  The modified constraint
(\ref{finalN}) is then enough to fix the correction to the scalars to be
\bea
&& M^I=\bar M^I\left[1-\fft{c_2}{288}F^2+\mathcal O(c_2^2)\right] \, .
\label{eq:mlo}
\ea
Consistency with the equations of motion will presumably demand an appropriate
relation between the various $c_{2I}$ coefficients.  However, since the
vectors will be truncated out, we only care about the combination $c_2$ given
in (\ref{eq:sans1}), and will not work out this relation explicitly.

We are now ready to integrate out both the scalars $M^I$ and the auxiliary
fields from the two-derivative action $\mathcal L_0$ given in (\ref{L0total}).
By making use of the corrections%
\footnote{These can be easily verified using $P_I = \frac{1}{4}
\bar{\mathcal N}_{IJ}\bar{M}^J$.}
to the leading order scalar expressions (\ref{globalAds})
\beq
P_I M^I = \frac{3}{2}\left[1-\frac{c_2}{288}F^2 \right] \, , \qquad
(\mathcal N^{-1})^{IJ}P_I P_J=\frac{3}{8}\left[1+\frac{c_2}{288}F^2\right]\, ,
\eq
we find that the contribution coming from ${\cal L}_0$ yields the following terms:
\begin{equation}
\label{L0final}
e^{-1}
{\cal L}_0=-R-\fft34F^2+\fft{1}{4}\epsilon^{\mu\nu\rho\lambda\sigma}
A_\mu F_{\nu\rho} F_{\lambda \sigma} + 12g^2
+ \fft{c_2}{24}\left[
\fft1{16}RF^2+\fft1{64}(F^2)^2 - \fft{5}{4}  g^2 F^2 \right]\, .
\end{equation}

Turning next to the four-derivative contributions, we note that, since
such terms are already linear in $c_2$, we may simply use the leading order
solution for the scalars.  The gauging contribution (\ref{L1gauged}) is then
particularly simple
\begin{equation}
e^{-1}{\cal L}_1^{\rm gauged}  = - \frac{c_2}{64} g^2\,
\epsilon_{\mu\nu\rho\lambda\sigma} A^\mu F^{\nu\rho}F^{\lambda\sigma} \, .
\end{equation}
On the other hand, the contribution to ${\cal L}_1^{\rm ungauged}$ is given by:
\bea
e^{-1}{\cal L}^{\rm ungauged}_1 &=& \frac{c_2}{24} \Bigl[
\frac{1}{16} \; \epsilon_{\mu \nu \rho \lambda \sigma} A^\mu
R^{\nu \rho \delta \gamma} R^{\lambda \sigma}{}_{\delta \gamma} +
\frac{1}{8} \, C_{\mu\nu\rho\sigma}^2
 + \frac{3}{16} C_{\mu \nu \rho \lambda}F^{\mu \nu} F^{\rho \lambda}
- F^{\mu \rho} F_{\rho \nu} R^{\nu}_{\mu} \nn \\
&&\qquad - \frac{1}{8} R F^2  + \frac{3}{2} \, F_{\mu \nu} \nabla^\nu
\nabla_\rho F^{\mu \rho} + \frac{3}{4} \, \nabla^\mu F^{\nu \rho}
\nabla_\mu F_{\nu \rho} +
\frac{3}{4} \, \nabla^\mu F^{\nu \rho} \nabla_\nu F_{\rho\mu} \nn \\
&&\qquad + \frac{1}{8} \; \epsilon_{\mu \nu \rho \lambda \sigma}
F^{\mu \nu} (3 F^{\rho \lambda} \nabla_\delta F^{\sigma \delta} +
4 F^{\rho \delta} \nabla_\delta F^{\lambda \sigma} + 6 F^{\rho}_{\;\;\delta} \nabla^\lambda F^{\sigma \delta})\nn \\
&&\qquad + \frac{45}{64} F_{\mu \nu} F^{\nu \rho}F_{\rho \lambda}
F^{\lambda\mu} - \frac{45}{256}(F^2)^2\Bigr] \, .
\ea
The full on-shell Lagrangian is thus given by
\bea
e^{-1}{\cal L} &=&-R-\fft34F^2 \Bigl(1 + \frac{5}{72} c_2g^2 \Bigr)
+\fft{1}{4}\Bigl(1-\fft1{16}c_2g^2\Bigr)\epsilon^{\mu\nu\rho\lambda\sigma}
A_\mu F_{\nu\rho} F_{\lambda \sigma} + 12g^2  \nn \\
&&+ \fft{c_2}{24}\Bigl[
\fft1{16}RF^2+\fft1{64}(F^2)^2\Bigr]
+ {\cal L}_1^{\rm ungauged}\, .
\ea
Finally, we may redefine $A_\mu$ to write the kinetic term in canonical form:
\beq
A_\mu^{\text{final}} = \sqrt{3}\Bigl(1+ \frac{5}{144}c_2g^2\Bigr)
\, A_{\mu}^{\text{old}} \, .
\eq
The Lagrangian then becomes:
\bea
\label{OurFinalL}
{\cal L} &=&-R-\fft14F^{2}
+\fft{1}{12\sqrt{3}}\Bigl(1-\frac{1}{6}c_2g^2\Bigr)
\epsilon^{\mu\nu\rho\lambda\sigma}
A_\mu F_{\nu\rho} F_{\lambda \sigma} + 12 g^2 \nn \\
&&+ \fft{c_2}{24}\Bigl[ \fft1{48}RF^2+\fft1{576}(F^2)^2\Bigr] +
{\cal L}_1^{\rm ungauged} \, ,
\ea
with
\bea
\label{L1ungaugedFINAL}
e^{-1}{\cal L}^{\rm ungauged}_1 \!&=& \frac{c_2}{24} \Bigl[
\frac{1}{16\sqrt{3}}\epsilon_{\mu \nu \rho \lambda \sigma} A^\mu
R^{\nu \rho \delta \gamma} R^{\lambda \sigma}{}_{\delta \gamma} +
\frac{1}{8}  C_{\mu\nu\rho\sigma}^2
+ \frac{1}{16} C_{\mu \nu \rho \lambda}F^{\mu \nu} F^{\rho \lambda}
- \frac{1}{3} F^{\mu \rho} F_{\rho \nu} R^{\nu}_{\mu} \nn \\
&&\qquad - \frac{1}{24} R F^2  + \frac{1}{2}  F_{\mu \nu} \nabla^\nu
\nabla_\rho F^{\mu \rho} + \frac{1}{4}  \nabla^\mu F^{\nu \rho}
\nabla_\mu F_{\nu \rho} +
\frac{1}{4}  \nabla^\mu F^{\nu \rho} \nabla_\nu F_{\rho\mu} \nn \\
&&\qquad + \frac{1}{32\sqrt{3}}  \epsilon_{\mu \nu \rho \lambda \sigma}
F^{\mu \nu} (3 F^{\rho \lambda} \nabla_\delta F^{\sigma \delta} +
4 F^{\rho \delta} \nabla_\delta F^{\lambda \sigma} + 6 F^{\rho}{}_{\delta} \nabla^\lambda F^{\sigma \delta})\nn \\
&&\qquad + \frac{5}{64} F_{\mu \nu} F^{\nu \rho}F_{\rho \lambda}
F^{\lambda\mu} - \frac{5}{256}(F^2)^2\Bigr] \, .
\ea
%

\section{Anomaly matching and AdS/CFT}

In the above section, we have written out the on-shell five-dimensional
$\mathcal N=2$ gauged supergravity Lagrangian up to four-derivative order.
Restoring Newton's constant, this takes the form
\begin{equation}
e^{-1}\mathcal L=\fft1{16\pi G_5}\left[-R-\fft14F^2+\fft1{12\sqrt3}
\epsilon^{\mu\nu\rho\lambda\sigma}A_\mu F_{\nu\rho}F_{\lambda\sigma}
+12g^2+\fft{c_2}{192}C_{\mu\nu\rho\sigma}^2+\cdots\right],
\label{eq:n=2sub}
\end{equation}
where we have only written out a few noteworthy terms.  Given this Lagrangian,
it is natural to make the appropriate AdS/CFT connection to $\mathcal N=1$
super-Yang Mills theory.  Before we do so, however, we present a brief
review of the AdS/CFT dictionary in the case of $\mathcal N=4$ super-Yang
Mills.

The standard AdS/CFT setup relates IIB string theory on AdS$_5\times S^5$
to $\mathcal N=4$ super-Yang Mills with gauge group SU($N$) and 't~Hooft
coupling $\lambda=g_{YM}^2N$.  The standard AdS/CFT dictionary then reads
\begin{equation}
\fft{L^4}{\alpha'^2}=4\pi g_sN=g_{YM}^2N,
\label{eq:adsdict}
\end{equation}
where $L$ is the `radius' of AdS$_5$.  This duality may be approached more
directly by reducing IIB supergravity on $S^5$, yielding $\mathcal N=8$
gauged supergravity in five dimensions.  Just as in the $\mathcal N=2$ case
of (\ref{eq:n=2sub}), this theory is determined in terms of two gravity-side
parameters, $G_5$ (Newton's constant) and $g$ (the gauged supergravity
coupling constant).  These are related to the parameters of the AdS/CFT
dictionary (\ref{eq:adsdict}) according to
\begin{equation}
g=\fft1L,\qquad N^2=\fft{\pi L^3}{2G_5}.
\label{eq:5ddict}
\end{equation}

Since the range of $\mathcal N=1$ gauge theories is much richer than that of
$\mathcal N=4$ SYM, it is worth rewriting the above AdS/CFT relations in
terms of more general invariants of the gauge theory.  This may be elegantly
done through anomaly matching, and in particular by making a connection through
the holographic Weyl anomaly \cite{Henningson:1998gx}.
Note that a discussion of the $\mathcal N=1$ SCFT description of
the higher derivative theory was already given in
\cite{Hanaki:2006pj}, where special emphasis was placed on the
technique of $a$-maximization.  Here we wish to
provide a more complete discussion of the relation between the
gravity parameters $G_5$, $g$ and $c_2$ and the gauge theory data.

\subsection{The Weyl anomaly}

For a four-dimensional field theory in a curved background, the Weyl anomaly
may be parameterized by two coefficients, commonly denoted $a$ and $c$ (or
equivalently $b$ and $b'$)
\begin{equation}
\langle T^\mu_\mu\rangle=\fft{c}{16\pi^2}\,C-\fft{a}{16\pi^2}E,
\label{eq:cftweyl}
\end{equation}
where
\begin{equation}
C=C_{\mu\nu\rho\sigma}^2=R_{\mu\nu\rho\sigma}^2-2R_{\mu\nu}^2+\ft13R^2
\end{equation}
is the square of the four-dimensional Weyl tensor, and
\begin{equation}
E=\widetilde R_{\mu\nu\rho\sigma}^2=R_{\mu\nu\rho\sigma}^2-4R_{\mu\nu}^2+R^2
\end{equation}
is the four-dimensional Euler invariant.  At the two-derivative level, the
holographic computation of the $\mathcal N=4$ SYM Weyl anomaly gives
$a=c=N^2/4$ \cite{Henningson:1998gx}.  Combining this with (\ref{eq:5ddict})
then allows us to write
\begin{equation}
a=c=\fft{\pi L^3}{8G_5},
\label{eq:acrel2d}
\end{equation}
which has the advantage of being completely general, independent of the
particular gauge theory dual.

The prescription for obtaining the holographic Weyl anomaly for higher
derivative gravity was worked out in \cite{Blau:1999vz,Nojiri:1999mh}, and
later extended in \cite{Fukuma:2001uf} for general curvature squared terms.
The result is that, for an action of the form
\begin{equation}
e^{-1} {\mathcal L} = \frac{1}{2\kappa^2} \left(-R+12g^2
+ \alpha R^2 + \beta R_{\mu \nu}^2 + \gamma R_{\mu \nu \rho \sigma}^2+\cdots
\right)\;,
\end{equation}
the holographic Weyl anomaly may be written as \cite{Fukuma:2001uf}
\begin{equation}
g_{\mu\nu}\langle T^{\mu\nu} \rangle
=\frac{2L}{16\pi G_5} \left[ \Bigl(- \frac{L}{24} + \frac{5\alpha}{3}
+ \frac{\beta}{3} + \frac{\gamma}{3} \Bigr) R^2
+ \Bigl( \frac{L}{8} - 5 \alpha - \beta
- \frac{3 \gamma}{2} \Bigr) R_{\mu \nu}^2 + \frac{\gamma}{2}
R_{\mu \nu \rho \sigma}^2 \right],
\label{eq:adsweyl}
\end{equation}
where $L$ is related to $g$ (to linear order) by
\begin{equation}
g=\fft1L\left[1-\fft1{6L^2}(20\alpha+4\beta+2\gamma)\right].
\end{equation}
Comparison of (\ref{eq:cftweyl}) with (\ref{eq:adsweyl}) then gives the
curvature-squared correction to (\ref{eq:acrel2d})
\begin{eqnarray}
a&=&\fft{\pi L^3}{8G_5}\left[1-\fft4{L^2}(10\alpha+2\beta+\gamma)\right]\nn\\
c&=&\fft{\pi L^3}{8G_5}\left[1-\fft4{L^2}(10\alpha+2\beta-\gamma)\right].
\end{eqnarray}

Turning now to the $\mathcal N=2$ gauged supergravity Lagrangian of
(\ref{eq:n=2sub}), we see that the curvature-squared corrections are
proportional to the square of the five-dimensional Weyl tensor.  This
gives
\begin{equation}
(\alpha, \beta, \gamma) = \frac{c_2}{192}
\left( \frac{1}{6}, - \frac{4}{3} , 1 \right),
\end{equation}
so that
\begin{equation}
a=\fft{\pi L^3}{8G_5},\qquad c=\fft{\pi L^3}{8G_5}
\left(1+\fft{c_2}{24L^2}\right),\qquad g=\fft1L.
\label{eq:weylrels}
\end{equation}
Note that the AdS radius is unshifted from that of the lowest order theory.
This is because AdS is conformally flat, so that the Weyl-squared
correction in (\ref{eq:n=2sub}) has no effect on the background.  Finally,
we may solve for $c_2$ to obtain
\begin{equation}
\fft{c_2}{24}=\fft{c-a}{ag^2}.
\label{c2trace}
\end{equation}
This is the key relation connecting the four-derivative terms in the
gauged supergravity Lagrangian to the $\mathcal N=1$ gauge theory data.

\subsection{The $R$-current anomaly}

A consistency check on the form of $c_2$ comes from the
gravitational contribution to the anomalous divergence of the
U(1)$_R$ current $\langle \p_\mu(\sqrt{g}\, \mathcal R^\mu)\rangle$, since
the latter is related by supersymmetry to the conformal anomaly $\langle T^\mu_\mu \rangle$.

The CFT $U(1)$ anomaly is given by
\begin{equation}
\delta_I (\Lambda) Z_{CFT} = \int \Lambda^I \left[ \frac{\mbox{tr} (G_I G_J G_K)}{24 \, \pi^2}
F^J \wedge F^K + \frac{\mbox{tr} \, G_I}{192 \,\pi^2} R_{ab} \wedge R^{ab} \right],
\end{equation}
where $G_I$ is a global $U(1)_I$ generator, and the trace is
taken to be a sum over all the fermion loops. The AdS/CFT
relation $Z_{CFT}=\exp(-I_{bulk})$ then connects this field theory anomaly to
the coefficients of the Chern-Simons terms in the bulk supergravity:
\begin{equation}
I_{bulk} = \cdots + \int \left[\frac{\mbox{tr} (G_I G_J G_K)}{24 \, \pi^2} A^I \wedge F^J \wedge F^K
+ \frac{\mbox{tr} \,G_I}{192 \, \pi^2} A^I \wedge R_{ab} \wedge R^{ab} \right]\;,
\label{Anomalies}
\end{equation}
where the ellipses denote the gauge invariant part of the action.
Comparison to the $A\wedge R\wedge R$ term of (\ref{ungauged}) gives
\begin{equation}
\mbox{tr} \, G_I = - \frac{\pi c_{2I}}{8G_5}.
\end{equation}
To relate $c_2\equiv c_{2I} \bar{M}^I$ to the central charges, we can use
the relation
\begin{equation}
a= \frac{3}{32} (3 \mbox{tr} R^3 - \mbox{tr} R),\quad c= \frac{1}{32} (9 \mbox{tr} R^3 - 5 \mbox{tr} R)\,,
\label{Centrals}
\end{equation}
provided we can relate $G_I$ appropriately to the U(1) charges $R$. A few
comments are needed to explain how to identify the $R$-charge
correctly. First of all, the $R$-charge is a particular linear
combination of the $G_I$, proportional to $\bar{M}^I G_I$.
Also, the supercharge $Q_{\alpha}$ should have $R$-charge one.
The U(1) charges of $Q_{\alpha}$ can be read off from the
coupling between the gauge fields and the graviphoton in the
gravity side, and the algebra is given by $[G_I, Q_{\alpha}] =
P_I Q_{\alpha}$. This uniquely determines the $R$-charge as
\begin{equation}
R = \frac{\bar{M}^I G_IL}{P_I \bar{M}^I}\quad \to \quad \mbox{tr} R
= - \frac1{P_I \bar{M}^I} \frac{\pi c_2L}{8G_5} \, .
\label{RSymm}
\end{equation}
Recall that the combination $P_I \bar{M}^I=3/2$ can be
determined from the vacuum solution,
(\ref{globalAds}). By plugging this equation into
(\ref{Centrals}), we obtain
\beq \frac{c_2}{24} = \fft{8G_5}{\pi L}(c-a)\, .
\label{CM1}
\eq

In addition, the gravitational constant also can be determined from the U(1)
anomaly. Eq.~(\ref{Anomalies}) implies
\begin{equation}
\mbox{tr} (G_I G_J G_K) = \frac{\pi}{8G_5}
\left(12 c_{IJK} - \fft{g^2}3 c_{(I} P_J P_{K)}\right)\, .
\end{equation}
By multiplying $\bar{M}^I \bar{M}^J \bar{M}^K$ on both sides,
we obtain
\begin{equation}
\frac{27}{8L^3} \mbox{tr} R^3 = \frac{\pi}{8G_5} \left(12 -\fft{3 c_2}{4L^2}\right)\, .
\end{equation}
The formula for the central charges (\ref{Centrals}) and
(\ref{CM1}) then gives
\begin{equation}
\frac{1}{16\pi G_5} = \frac{a}{2 \pi^2L^3} \,.
\label{CM2}
\end{equation}
Using this relation, (\ref{CM1}) can be rewritten as
\begin{equation}
\fft{c_2}{24L^2} = \frac{c - a}{a} \, .
\label{c2kentaro}
\end{equation}
These results agree with those found through the holographic Weyl anomaly
calculations, as expected for consistency.

\subsubsection{Extracting the R-current anomaly from the $\mathcal N =2$ case}

Since the U(1) normalization may be somewhat obscure, we may perform an
additional check by making contact with the ${\cal N}=2$ SCFT literature.
In fact, one can extract the $c_2$ result (\ref{c2trace}) from the analysis
of \cite{Aharony:1999rz}, which studied $R$-symmetry anomalies in
the ${\cal N}=2$ SCFT dual to AdS$_5 \times S^5/\mathbb{Z}_2$.
Of course, the appropriate supersymmetric CFT that is dual to our bulk
${\cal N}=2$ AdS$_5$ theory has ${\cal N}=1$ supersymmetry. Nevertheless, one
can still use the analysis of \cite{Aharony:1999rz}, after
carefully rewriting it in the language of ${\cal N}=1$
anomalies. Before doing so, we will need to make a few general
comments on the connection between the CFT $R$-current anomalies
and the dual supergravity description.

The four-dimensional CFT $R$-current anomaly is sensitive to the
amount of supersymmetry, and is given by \cite{Anselmi:1998zb}:
\bea \label{RanomN1}
 \p_\mu (\sqrt{g} \,{\cal R}^\mu )_{{\cal N}=1}  &=& \frac{c-a}{12 \pi^2} \, \tilde{R} R
+  \frac{5a-3c}{9 \pi^2} \, \tilde{F} F \, , \\
\label{RanomN2} \p_\mu (\sqrt{g} \,{\cal R}^\mu )_{{\cal N}=2}
&=& \frac{c-a}{4 \pi^2} \, \tilde{R} R +  \frac{3(c-a)}{\pi^2}
\, \tilde{F} F \, ,
\ea
where $F$ is the flux associated with
the $R$-symmetry. The $R$-symmetry of ${\cal N}=2$ SCFTs is
U(1)$_R\times \mathrm{SU}(2)_R$. The U(1)$_R$ symmetry of its
${\cal N}=1$ subalgebra is
\beq {\cal R}_{{\cal N}=1} = \frac{1}{3}
{\cal R}_{{\cal N}=2} + \frac{4}{3} I_3,
\eq
where
$I_1,I_2,I_3$ are $SU(2)_R$ generators. The factor of $1/3$ in
the relation above can also be seen in the gravitational
contributions to $\p_\mu (\sqrt{g} \,{\cal R}^\mu )$ in
(\ref{RanomN1}) and (\ref{RanomN2}). Recall that the mixed
U(1)-gravity-gravity anomaly $\p_\mu (\sqrt{g} \,{\cal R}^\mu
) \propto  \tilde{R} R$ is represented in the bulk by the mixed
gauge-gravity Chern-Simons interaction $\propto \int_{\rm AdS_5} A
\wedge \mbox{tr}(R \wedge R)$. Thus, the bulk CS term associated to
the ${\cal N}=1$ SCFT will be $1/3$ of that corresponding to
${\cal N}=2$.

Furthermore, when using the results of \cite{Aharony:1999rz},
we will have to be careful with how the U(1) gauge field is
normalized.
In the AdS/CFT dictionary, the normalization of the gauge field
kinetic term \beq S_{AdS_5}=\int d^4x \, dz \, \sqrt{-g} \;
\frac{F_{\mu\nu}F^{\mu\nu}}{4\, g_{SG}^2} \eq can be extracted
by looking at the two-point function of the dual CFT currents
sourced by the gauge field $A_\mu (\vec{x})=A_\mu
(\vec{x},z)|_{\rm boundary}$. For a four-dimensional CFT, the
general form of the two point function of such currents is
given by \cite{Freedman:1998tz}: \beq \langle J_i(x) J_j(y)
\rangle = \frac{B}{(2\pi)^4} \bigl(\,\Box \delta_{ij} -
\p_i\p_j\bigr) \frac{1}{(x-y)^4} \, , \eq where $B$ is a
numerical coefficient which is related to the normalization of
the gauge kinetic term: \beq B \propto \frac{1}{g_{SG}^2} \, .
\eq For the ${\cal N}=2$ computation of \cite{Aharony:1999rz}
one finds $B=8$, while for the case of ${\cal N}=1$
supersymmetry \cite{Anselmi:1996dd} we read off $B=8/3$. Notice
that the two results are again off by a factor of $3$. We now
have all the ingredients we need to apply the (${\cal N}=2$
SCFT) analysis of \cite{Aharony:1999rz} to our case (${\cal
N}=1$ SCFT). We have seen that both the gauge kinetic term
normalization and the coefficient of the mixed gauge-gravity CS
term will have to be adjusted.

The five-dimensional supergravity action of \cite{Aharony:1999rz}
takes the form
\bea
S &=& \frac{N^2}{\pi^2L^3} \int \sqrt{-g}
\; \frac{F_{\cal R}^2}{4} +
\frac{N}{16\pi^2L} \int  A^{\cal R} \wedge \mbox{tr} (R\wedge R) \nn \\
& = &  \frac{N^2}{4 \pi^2L^3} \int \Bigl[ \sqrt{-g} \; F_{\cal
R}^2 - \frac{L^2}{16 \, N} \, \epsilon_{\mu\nu\rho\lambda\sigma}
A^\mu R^{\nu \rho \delta \gamma}R^{\lambda \sigma}_{\;\;\;\;
\delta\gamma} \Bigr] \, ,
\ea
where $A^{\cal R}$ is the gauge
field that couples canonically to the $R$-current. This
was the effective supergravity Lagrangian which was appropriate for
comparison to the ${\cal N}=2$ SCFT. Since we are interested in
comparing to a CFT with ${\cal N}=1$ SUSY, we will need to
rescale both terms by appropriate factors of $1/3$:
\bea S
\rightarrow \frac{N^2}{4 \pi^2L^3} \int \Bigl[ \sqrt{-g}  \; \;
\frac{1}{3}\,F_{\cal R}^2 - \frac{L^2}{3\cdot 16 \, N} \,
\epsilon_{\mu\nu\rho\lambda\sigma} A^\mu R^{\nu \rho \delta
\gamma}R^{\lambda \sigma}{}_{\delta\gamma} \Bigr] \, .
\ea
Finally, we rescale the graviphoton, $A^{\cal
R}=(\sqrt{3}/{2})A$, to obtain a canonical gauge kinetic
term:
\beq S \rightarrow \frac{N^2}{4 \pi^2L^3} \int \Bigl[
\sqrt{-g}  \; \; \frac{F^2}{4} - \frac{L^2}{32 \sqrt{3} \, N} \,
\epsilon_{\mu\nu\rho\lambda\sigma} A^\mu R^{\nu \rho \delta
\gamma}R^{\lambda \sigma}{}_{\delta\gamma} \Bigr] \, .
\eq
This is the action which should be compared to ours:
\beq
S_{us}=\frac{N^2}{4\pi^2L^3} \int \sqrt{g} \Bigl[-R-\frac{F^2}{4}
+ \frac{c_2}{24 \cdot 16 \sqrt{3}} \,
\epsilon_{\mu\nu\rho\lambda\sigma} \, A^\mu R^{\nu \rho \delta
\gamma}R^{\lambda \sigma}_{\;\;\;\;
\delta\gamma}+\ldots\Bigr]\, ,
\eq
finally giving us
\beq
c_2 =\frac{12L^2}{N} =24L^2 \frac{c-a}{a}\, ,
\eq
in agreement with (\ref{c2trace}) and (\ref{c2kentaro}).

\section{$R$-Charged Solutions}

The embedding of the lowest order five-dimensional ${\cal N}=2$ gauged $U(1)^3$
supergravity into IIB supergravity was done in \cite{Cvetic:1999xp}.
If the three $U(1)$ charges are taken to be equal, we end up with the
minimal supergravity system that we have considered above, (\ref{eq:l0os}).
The static stationary non-extremal solutions are well known, and were found
in \cite{Behrndt:1998jd}.  For the truncation to minimal supergravity, they
take the form
\bea
\label{0sol}
ds^2&=& H^{-2}fdt^2-H\Big(f^{-1}dr^2+r^2d\Omega_{3,k}^2 \Big), \nonumber \\
A &=& \sqrt{\frac{3(kQ+\mu)}{Q}}\Big(1-\frac{1}{H}\Big)dt,
\ea
where the metric functions $H$ and $f$ are:
\bea
\label{0sol1} H(r) &=& 1 + \frac{Q}{r^2}, \nonumber \\
f(r) &=& k -\frac{\mu}{r^2} + g^2r^2H^3 \, .
\ea
Here $\mu$ is a non-extremality parameter and $d\Omega^2_{3,k}$
for $k =$ $1$, $0$, or $-1$ corresponds to the unit metric of a spherical,
flat, or hyperbolic 3-dimensional geometry, respectively.

\subsection{Higher order corrected $R$-charged Solutions}

We wish to find corrections to the $R$-charged solutions (\ref{0sol})
given the higher derivative Lagrangian (\ref{OurFinalL}). To this
end, as in \cite{Liu:2008kt} we treat $c_2$ as a small parameter and expand the metric and
gauge field as follows:
\bea
\label{1sol}
H(r) &=& 1 + \frac{Q}{r^2} + c_2 h_1(r) \, ,\nonumber \\
f(r) &=& k -\frac{\mu}{r^2} + g^2r^2H^3 + c_2 f_1(r)\, , \nonumber \\
A &=& \sqrt{\frac{3(kQ+\mu)}{Q}}\Big(1-\frac{1+c_2 a_1(r)}{H}\Big)dt \, ,
\ea
where $h_1, f_1,$ and $a_1$ parameterize the corrections to the background geometry.
Solving the equations of motion for the theory, we arrive at:
\bea
h_1 &=& -\frac{Q(kQ+\mu)}{72r^6H_0^2} \, ,\nn \\
f_1 &=& \frac{-5 g^2Q(kQ+\mu)}{72r^4} + \frac{\mu^2}{96r^6H_0}\, ,\nn \\
a_1 &=& \frac{Q}{144r^6H_0^3}\big[4(kQ+\mu) - 3\mu -
\frac{3Q\mu}{r^2}\big] \, .
\ea
The new corrected geometry is therefore given by
\bea
H(r)&=& H_0(r)+ \frac{c_2}{24} \bigg[ \, \frac{-Q(kQ+\mu)}{3 r^6H_0^2} \bigg] \, , \nn\\
f(r) &=& f_0(r) +\frac{c_2}{24} \bigg[ - \frac{8g^2Q(kQ+\mu)}{3r^4} + \frac{\mu^2}{4r^6H_0} \bigg] \, ,\nn \\
A_t(r)&=& A_{t\,0}(r)-\frac{c_2}{24}
\frac{\sqrt{3Q(kQ+\mu)}}{2r^8H_0^4}\bigg[2(kQ+\mu)r^2-\mu r^2H_0\bigg]\, ,
\ea
where $H_0, f_0,$ and $A_0$ refer to the background solutions (\ref{0sol}) and (\ref{0sol1}).
Finally, we should note that in the literature $Q$ and $\mu$ are sometimes
written in terms of a parameter $\beta$, defined by
$\sinh^2 \beta = kQ/\mu^2$.

We will state the $k=0$ and $k=1$ solutions explicitly, since they have
several interesting applications: the former to studies of the hydrodynamic
regime of the theory, and the latter to the issue of horizon formation for
small black holes.
For $k=0$, the solution is given by
\bea
H(r)&=& H_0(r)+\frac{c_2}{24} \bigg[ \, \frac{-Q\mu}{3 \, r^6H_0^2} \bigg]
\, , \nn\\
f(r) &=& f_0(r) +\frac{c_2}{24} \bigg[ - \frac{8g^2\mu Q}{3r^4} +
\frac{\mu^2}{4 \, r^6H_0}\bigg]  \, , \nn\\
A_t(r)&=& A_{t\,0}(r)- \frac{c_2}{24}
\bigg[\frac{\sqrt{3Q\mu}}{2r^8H_0^4}(\mu r^2 - Q\mu) \bigg] \, .
\label{HfAcorrected}
\ea
while for $k=1$ it is given by
\bea
\label{k1corrected}
H(r)&=& H_0(r) - \frac{c_2}{24} \bigg[ \, \frac{Q(Q+\mu)}{3 r^2(r^2+Q)^2} \bigg] \, , \nonumber\\
f(r) &=& f_0(r) +\frac{c_2}{24} \bigg[ - \frac{8g^2Q(Q+\mu)}{3r^4} + \frac{\mu^2}{4r^6H_0} \bigg]  \, , \nonumber \\
A_t(r)&=& A_{t\,0}(r)-\frac{c_2}{24}\bigg[
\frac{\sqrt{3Q(Q+\mu)}}{2r^8H_0^4}\Big((2Q+\mu) r^2 -Q\mu\Big)\bigg]\,.
\ea

\subsection{Conditions for Horizon Formation}

We would like to conclude this section with some comments on the structure of
the horizon for the solutions that we have found.
In particular, we are interested in whether higher derivative
corrections will facilitate or hinder the formation of a horizon.
In the standard two-derivative theory, the BPS-saturated limit
($\mu=0$) of the $k=1$ solution (\ref{0sol})-(\ref{0sol1}) describes
a geometry with a naked singularity, the so-called superstar
\cite{Myers:2001aq}.
Furthermore, even if the non-extremality parameter is turned on, one
finds that a horizon develops only given a certain critical amount,
$\mu \geq \mu_c$ \cite{Behrndt:1998jd}. It is therefore natural to
ask what happens to such geometries once we start incorporating
curvature corrections. For the superstar, we would like to see hints
of horizon formation. In the non-extremal case, on the other hand,
it would be nice to determine whether the inclusion of higher-derivative
corrections leads to a smaller (larger) critical value $\mu_c$,
increasing (decreasing) the parameter space for the appearance of a
horizon. However, one should keep in mind that our arguments are
only suggestive, since our analysis is perturbative, while the
formation of a horizon is a non-perturbative process.
Moreover, given that even in the non-extremal case turning on $\mu$ does not
guarantee the presence of a horizon, it is not clear at all whether
higher derivative corrections can be enough to push the superstar to develop
a horizon.
A more proper analysis would involve looking directly at the SUSY conditions,
and asking whether they are compatible with having a superstar solution with a finite
horizon. In fact,
there are already studies which seem to indicate \cite{Mueck} that
this may not be possible.

The spherically symmetric solutions presented in (\ref{k1corrected})
are of the form:
\beq
ds^2 = F_1(r) \,dt^2 - F_2(r) \,dr^2 - F_3(r) \, d\Omega_3^2 \, .
\eq
Horizons appear at zeroes of the function $F_1(r)$.
One can make arguments about their existence without having to solve
explicitly for their exact location.
Notice that $F_1(r)$ is a positive function for large $r$.
Thus, a sufficient condition for having at least
one horizon is
\beq
F_1(r_{min})\leq 0 \, ,
\eq
where $r_{min}$ is a (positive) minimum of $F_1(r)$.
This was the reasoning used in \cite{Behrndt:1998jd} to study
the properties of the horizon of the non-extremal solution.

For the corrected superstar solution we have, expanding in $c_2$:
\beq F_1 \equiv \frac{f}{H^2}  = \frac{f_0+c_2(f_1-2f_0 h_1
H_0^{-1})}{H_0^2} + \mathcal O(c_2^2) \, . \eq It is easy to see
that, to leading order, the numerator does not vanish. With the
inclusion of higher-derivative terms, however, it picks up a
negative contribution, hinting at the possibility of a horizon.
Furthermore, the minimum of the function $F \equiv f_0+c_2(f_1-2f_0
h_1 H_0^{-1})$ will shift. Let's see precisely how that happens. To
lowest order, its minimum is given by $x_{min}^{(0)}=2Q$, which in
turn gives us $F(x_{min}^{(0)})=1+27 g^2 Q/4$. Including higher
order corrections, we find
\beq
x_{min}=x_{min}^{(0)}+c_2
x_{min}^{(1)} = 2Q - c_2 \frac{81 g^2 Q -4}{4374 Q g^2}\, .
\eq
Now we have $$ F(x_{min})= 1+27 g^2 Q/4 +
c_2(\frac{1}{972Q}-\frac{g^2}{48}),$$ which tells us that the
minimum of the function will be slightly closer to zero as long as
$g^2 Q > 4/81$.

The analysis of the conditions for the existence of a horizon
in the non-extremal case ($\mu \neq 0$) is significantly more involved.
The expression for the corrected horizon radius in terms of
the original, two-derivative horizon radius $r_0$ is:
\begin{eqnarray}
r_H &=& r_0\bigg( 1 +
\frac{c_2}{24}\Big\{\frac{g^4H_0^4(3Q^2-26Qr_0^2+3r_0^4) -
2g^2H_0^2(13Q-3r_0^2) + 3}{24H_0r_0[g^2H_0^2(Q-2r_0^2)-1]}
\Big\}\bigg) \, .
\end{eqnarray}
Notice that we traded $\mu$ in favor or $r_0$ in the expression above by making use
of $f_0(r_0)=0$, \emph{i.e.} the relation $\mu/r_0^2 = 1 + g^2r_0^2H_0^3$.
As we mentioned above, in the two-derivative case one finds a
critical value $\mu_{crit}$ above which a horizon will form.
It would certainly be interesting to explore for which
parameter values $r_H$ decreases or increases, and more importantly,
how the (corrected) critical value of $\mu$ is affected by the
curvature corrections. We leave this to future studies.

\section{Thermodynamics}

We may now study some of the basic thermodynamic properties of the non-extremal
solutions constructed above.  With an eye towards AdS/CFT in the Poincar\'e
patch, we will focus on the $k=0$ solution (\ref{HfAcorrected}), although the
analysis may easily be carried out for the other cases as well.  We begin with
the entropy, which for Einstein gravity is characterized by the area of the
event horizon. In the presence of higher derivative terms, however, this
relation is modified, and the entropy is no longer given by the area law.
Instead, we may turn to the Noether charge method developed in
\cite{Wald:1993nt} (see also \cite{Jacobson:1993vj, Iyer:1994ys}).

The original Noether charge method is only applicable to a theory with general
covariance, but has been extended to a theory with gravitational Chern-Simons
terms in \cite{Tachikawa:2006sz}.
Our action includes a mixed Chern-Simons term of the form $A \wedge R \wedge R$.
But as long as we keep this term as it is, with a bare gauge potential, the
general covariance is unbroken and we can still use the original formulation.
In the absence of covariant derivatives of the Riemann tensor, the entropy formula is given by \cite{Wald:1993nt}
\begin{equation}
S = - 2 \pi \int_{\Sigma} d^{D-2} x \sqrt{- h} \frac{\delta {\mathcal L}}{\delta R_{\mu \nu \rho \sigma}}
\, \epsilon_{\mu \nu} \epsilon_{\rho \sigma} \, ,
\label{entropy}
\end{equation}
where $\Sigma$ denotes the horizon cross section, $h$ is the induced metric
on the it and $\epsilon_{\mu \nu}$ is the binormal to the horizon cross
section.

For the metric ansatz (\ref{0sol}) the only non-vanishing component of the
binormal $\epsilon_{\mu\nu}$ is
\begin{equation}
\epsilon_{tr} = - \epsilon_{rt} = H^{-1/2} \, .
\end{equation}
Applying the prescription (\ref{entropy}) to the action (\ref{OurFinalL}),
we obtain, to linear order in $c_2$,
\begin{eqnarray}
S&=& \frac{A}{8G_5} \biggl[ - g^{\mu \rho} g^{\nu \sigma} + \fft{c_2}{24}
\bigl( -\ft14 C^{\mu \nu \rho \sigma} - \ft1{32} g^{\mu \rho} g^{\nu \sigma} F^2 + \ft5{12}
g^{\nu \sigma} F^{\mu \lambda} F^{\rho}{}_\lambda
- \ft1{16} F^{\mu \nu} F^{\rho \sigma}\bigr)\biggr] \epsilon_{\mu \nu} \epsilon_{\rho \sigma}\bigg|_{r = r_+}\nonumber\\
&=& \frac{A}{4G_5} \biggl[1 + c_2 \, \frac{\mu (Q +
3r_0^2)}{48(r_0^2 + Q)^3} \biggr]\, ,
\label{preentropy}
\end{eqnarray}
where $A = \int \sqrt{-h} \; d\Omega_{3,0}$ is the area of the horizon for
the solution to the higher derivative theory.  Also, $r_+$ denotes the radius
of the event horizon for the corrected black brane solution,
while $r_0$ is the horizon location for the original, two-derivative solution
(\ref{0sol1}).  The former can be found by requiring that the
$g_{tt}=f(r)/H(r)^2$ component of the corrected metric vanishes%
\footnote{To linear order in the expansion parameter $c_2$,
this coincides with demanding that $f(r)$ vanishes.}.
Similarly, $r_0$ satisfies $f_0(r_0)=0$.
Notice that the non-extremality parameter $\mu$ can be expressed
entirely in terms of $r_0$ and $Q$:
\beq f_0(r_0)=0 \;\; \Rightarrow
\;\; \mu = \fft{g^2(r_0^2 + Q)^3}{r_0^2}\,.
\eq
We can therefore
eliminate $\mu$ from (\ref{preentropy}), and write the entropy in
the following form:
\beq S=\frac{A}{4G_5} \Bigl[1 + c_2g^2
\frac{Q + 3r_0^2}{48\,r_0^2} \Bigr]\, .
\label{eq:SA}
\eq
The first term above is simply the contribution coming from the area, while
the remaining $\mathcal O (c_2)$ term is the expected deviation from the area
law.

In order to arrive at the entropy density, we need one more ingredient, which
is the relation between the corrected and uncorrected horizon radii $r_+$ and
$r_0$:
\begin{equation}
r_+ = r_0 \left( 1 + \frac{c_2}{24} \frac{g^2 (r_0^2 + Q) (3 Q^2 -
26 Q r_0^2 + 3 r_0^4)}{24 r_0^4 (Q - 2r_0^2)} \right) \, .
\label{radius}
\end{equation}
This is because the area $A$ appearing in (\ref{eq:SA}) is computed using
$r_+$.  This expression allows us to write the entropy per unit
three-brane spatial volume entirely in terms of $r_0$ as well as the
physical parameters of the theory
%
%
\begin{eqnarray}
\label{entropydens}
s &=& \frac{(r_0^2 + Q)^{3/2}}{4G_5L^3} \left(1+
\frac{c_2}{24} \frac{g^2(3Q^2 - 14 Q r_0^2 - 21 r_0^4)}{8r_0^2 (Q -
2 r_0^2)}\right)\nn\\
&=&\frac{2(r_0^2 + Q)^{3/2}}{\pi L^6} \left( a + (c-a)  \frac{3Q^2 - 14
Q r_0^2 - 21 r_0^4}{8 r_0^2 (Q - 2 r_0^2)}\right) \, .
\end{eqnarray}
In the second line we have used the relations (\ref{eq:weylrels}) to
replace the gravitational quantities $G_5$ and $c_2$ by the central charges
of the dual CFT.
Notice that the lowest order term above matches the two-derivative entropy computation of \cite{Son:2006em}.

While $r_0$ is the coordinate location of the horizon in the lowest order
computation, it is not in itself a physically relevant parameter.  Instead,
it may be viewed as a proxy for the Hawking temperature associated with the
non-extremal solution.  A simple way of computing this temperature is to
identify it with the inverse of the periodicity of Euclidean time $\tau$.
The relevant components of the metric are given by
\begin{equation}
ds^2=H^{-2}f d\tau^2+Hf^{-1}dr^2+\cdots,
\end{equation}
and the horizon is located at $f(r_+)=0$.  Expanding near the horizon and
identifying the proper period of $\tau$ to remove the conical singularity
yields the temperature
\begin{equation}
T_H = \frac{(r_0^2 + Q)^{1/2}}{2 \pi L^2} \Bigl[\frac{ (2 r_0^2 -
Q)}{r_0^2} + \frac{c_2}{24L^2} \frac{(3 Q^3 + 4 Q^2
r_0^2 + 59 Q r_0^4 - 10 r_0^6)}{8 r_0^4(2 r_0^2 - Q)}\Bigr] \, .
\end{equation}
In principle, we may invert this expression to obtain $r_0$ as a function
of temperature $T_H$ and charge $Q$.  This then allows us to rewrite
the entropy density as a function of charge and temperature,
$s=s(T_H,Q)$.  In practice, however, non-trivial $R$-charge introduces
a new scale, so that the entropy density/temperature relation no longer
takes the simple form $s\sim T^3$ resulting from simple dimensional analysis.

\section{Discussion}

The main result of the previous section is the derivation of the
entropy (\ref{entropydens}) of an $R$-charged black brane including
higher-derivative corrections, which are controlled by the parameter $c_2$.
Furthermore, the identification of the gravitational parameters
$G_5$ and $c_{2}$
in terms of the central charges $a$, $c$ of the dual CFT has allowed us to
express the entropy in terms of microscopic, gauge theory data.
In particular, the relation between $c_2$, which signals the
contribution coming from $R^2$ terms, and the CFT central charges
is given by $c_2 = 24L^2(c-a)/a$.

A non-trivial check on the corrections to the entropy
can be done by considering the zero $R$-charge ($Q=0$)
limit of (\ref{entropydens}),
which should agree with the analysis of \cite{Kats:2007mq}.
For a Lagrangian of the form
\beq
\label{Lkp}
\mathcal L = \frac{R}{16\pi G_5} - \Lambda + \alpha_1 R^2 + \alpha_2 R_{\mu\nu}^2 +
\alpha_3 R_{\mu\nu\rho\sigma}^2 \, ,
\eq
the authors of \cite{Kats:2007mq} showed that the entropy density of a 5D AdS black brane solution is given by
\beq
s = \frac{2\pi}{L^3 z_0^3}\Bigl[\frac{1}{8\pi G_5} - \frac{18}{L^2}(5\alpha_1+\alpha_2)
+ \frac{12}{L^2}\alpha_3 \Bigr] \, ,
\eq
where $L$ denotes the AdS curvature radius, $L^2 = -{6}/({8\pi G_5 \Lambda})$.
Comparing (\ref{Lkp}) to our action, where the only curvature corrections that
survive the $Q=0$ limit come in the form of
$C_{\mu\nu\rho\sigma}^2=\ft16 R^2 - \ft43 R_{\mu\nu}^2+R_{\mu\nu\rho\sigma}^2$,
we read off:
\bea
16\pi G_5 \alpha_1 = \frac{1}{48} \frac{c_2}{24}\, , \quad
16\pi G_5\alpha_2 = - \frac{1}{6} \frac{c_2}{24}\, , \quad
16\pi G_5\alpha_3 = \frac{1}{8} \frac{c_2}{24}\, .
\ea
Making use of these expressions, the entropy of \cite{Kats:2007mq} takes the form
\beq
s = \frac{1}{4L^3 z_0^3 G_5}\Bigl[1+\frac{21}{16}\frac{c-a}{c} \Bigr] \, ,
\eq
matching nicely the $Q=0$ limit of (\ref{entropydens}), as expected.

We should point out that a similar discussion has appeared very recently
in \cite{Buchel:2008vz}, where the authors considered the hydrodynamic regime
of the CFT dual to the zero $R$-charge black brane background of
\cite{Kats:2007mq}.
In \cite{Buchel:2008vz}, however, higher derivative corrections
associated with $R^2$ and $R_{\mu\nu}$ are eliminated via a field
redefinition, making direct comparison to our entropy less
straightforward.

Our interest in studying higher order corrections to $R$-charged
$AdS_5$ black holes is also motivated by our desire to investigate
corrections to the hydrodynamic regime of the dual theory.
It is natural to apply the results of this work to
the calculation of $\eta/s$, the shear viscosity to entropy ratio,
which has recently received a great deal of attention.
In particular, our present construction of higher-derivative
corrected $R$-charged black holes allows for a generalization of
the finite coupling shear viscosity calculation to the case of
finite ($R$-charge) chemical potential.
This is an avenue which we are currently exploring \cite{Cremonini:2009sy}.

We would like to conclude with a few comments on the issue of
horizon formation. As we mentioned in section III, the so-called
superstar solution at the two-derivative level has a naked
singularity. With the inclusion of higher derivative contributions,
it appears that the corrected superstar may develop a horizon,
provided that the charges are large enough, $g^2 Q >4/81$.
However, we should note that our analysis is entirely
perturbative, while horizon formation is an intrinsically
non-perturbative phenomenon.
While our results show that the first corrections to the geometry
seem to push the superstar solution ``in the right direction," increasing the
chances of forming a horizon, a more rigorous analysis is certainly needed to reach
a conclusive result.

\begin{acknowledgments}
We would like to thank A. Castro, R. Myers, A. Sinha and Y. Tachikawa for
useful comments and clarifications.  We would especially like to thank
A. Buchel for many valuable discussions and input.  S.C. is grateful for the
hospitality of Perimeter Institute, where part of this work was completed.
This work is supported in part by the US Department of Energy under grant
DE-FG02-95ER40899.
\end{acknowledgments}


\end{document}